\documentclass[aps,prl,twocolumn,superscriptaddress,showpacs]{revtex4}
\usepackage{epsfig}
\begin{document}
\title{Study of
the Localization Transition on a Cayley-tree via Spectral Statistics}
\author{Miri Sade and Richard Berkovits}
\affiliation{The Minerva Center, Department of Physics,
    Bar-Ilan University, Ramat-Gan 52900, Israel}
\date{March 21, 2003, version 1.0}
\begin{abstract}
The spectral statistics of a Cayley-tree is numerically studied.
The statistics are non-universal due to the high ratio of boundary
sites. Once the boundary sites are connected to each other in a
way that preserves the local structure of the tree the universal
statistics of the spectra is recovered. A clear localization
transition is observed as function of on-site disorder strength,
with a critical disorder $W_c=11.44^{+0.08}_{-0.04}$ and critical
index $\nu=0.51^{+0.05}_{-0.04}$. The value of $\nu$ fits nicely
to its mean field value, while the value of $W_c$ is puzzling.
\end{abstract}
\pacs{ }

\maketitle

The properties of the Anderson transition have generated much
interest since it was first predicted \cite{rev}. The transition is
characterized by a lower critical dimension, believed to be equal
to two, below which all states are localized for any amount of
disorder. Above the lower critical dimension a transition between
extended and localized states appears at some critical value of
disorder (or energy). An upper critical dimension, above which the
transition may be described by a mean field theory, is not well
established.

The Anderson transition is usually characterized by two
parameters. The critical disorder $W_c$ at the middle of the band
(where $W$ is the width of the distribution from which the on-site
energies are drawn in the canonical Anderson model defined in Eq.
(\ref{hamiltonian})) and the critical index $\nu$. The dependence
of $W_c$ and $\nu$ on the dimensionality $d$ has been the subject
of many recent numerical studies. For $d=3,4$ the values are well
established - $W_c \sim 16.5$, $\nu \sim 1.5$ for $d=3$
\cite{kramer93,slevin00} and $W_c \sim 35$, $\nu \sim 1$ for $d=4$
\cite{schreiber96,zharekeshev98}. The mean field value of the
critical exponent is equal to $1/2$. Assuming that the upper
critical dimension is equal to infinity, an extrapolation equation
for $\nu \sim 0.8/(d-2) + .5$ was proposed \cite{schreiber96}.
Verification of this extrapolation was obtained by studying the
Anderson transition for bifractal system, where it was
demonstrated that $d$ should be replaced by the spectral dimension
$d_s$. Similarly, the critical disorder is also extrapolated by
$W_c \sim 16.5(d-2)$ (again for bifractals $d$ is replaced by
$d_s$) \cite{schreiber96}.

Our main goal in this study is to identify the metal-insulator
transition in a disordered Cayley-tree, and to study its
properties numerically by spectral statistics. Although many
studies were performed for Cayley-tree structures, to the best of
our knowledge, no studies were performed using spectral statistics
\cite{shklov92}. Moreover, it is not trivial to extend the above
extrapolations for $\nu$ and $W_c$ to the Cayley-tree. From
analytical calculations it is known that for the Cayley-tree $\nu=0.5$
\cite{zirnbauer86,efetov87,mirlin91}, i.e., $d=\infty$. On the
other hand, a mobility edge is predicted at some finite disorder,
which is hard to conciliate with the extrapolation formula for
$W_c$ which gives $W_c=\infty$ if $d=\infty$ is plugged in, and
$W_c<0$ if the spectral dimension of
 a Cayley-tree $d_s=4/3$ \cite{havlin} is inserted.

We based our calculations on the usual tight-binding Hamiltonian,
\begin{equation}
   H=\sum_{i}\varepsilon_i a_i^{\dag} a_i -
   \sum_{<i,j>} a_j^{\dag} a_i , \label{hamiltonian}
\end{equation}
where the left part of $H$ stands for the disordered on-site
potential. The on-site energies, $\varepsilon_i$ are uniformly
distributed over the range $-W/2\leq\varepsilon_i\leq W/2 $. The
right part is the hopping element which is set to $1$, and $<i,j>$
denotes nearest neighbors. Here we considered a tree where each
site is connected to two sites below it. We diagonalize the
Hamiltonian exactly, and obtain $N$ eigenvalues $E_i$ (where $N$
is the number of sites in the tree) and eigenvectors $\psi_i$. The
calculations are made for $K$ different realizations -
$K=4000,2000,1000,\ldots, 125,64$ for the corresponding tree
sizes: $N=63,127,255,\ldots,2047,4095$ or $L=6,7,8,\ldots,11,12$
(where $L$ is the number of "generations" in the tree).

\begin{figure}\centering
\epsfxsize5.5cm\epsfbox{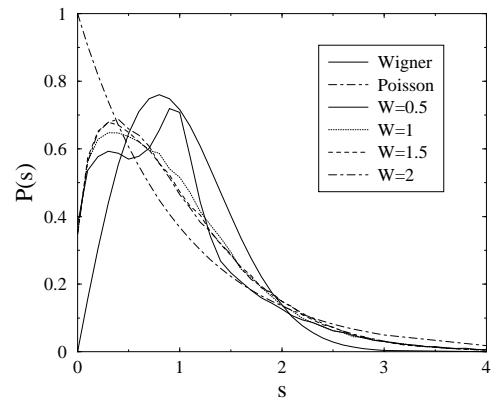} \caption{The distribution
$P(s)$ for an $L=7$ Cayley-tree for different $W$ values, with all
eigenvalues taken into account. The Wigner distribution (Eq.
(\ref{wigner}) as well as the Poisson distribution (\ref{poisson})
are indicated in the plot.} \label{fig1}
\end{figure}

\begin{figure}\centering
\epsfxsize5.5cm\epsfbox{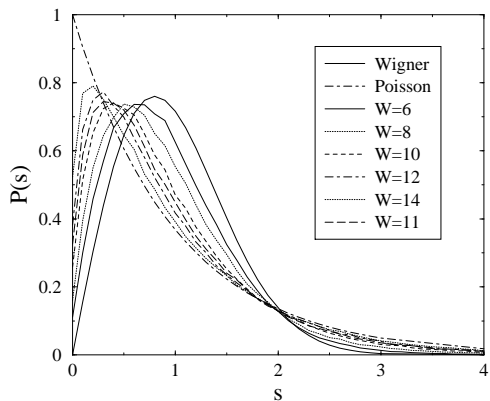} \caption{The distribution
$P(s)$ for an $L=10$ Cayley-tree in which the boundary leaves were
connected to each other, thus preserving the local Cayley-tree
structure while avoiding the peculiarities introduced by the
boundaries. A clear transition from Wigner to Poisson is observed
as function of disorder. } \label{fig2}
\end{figure}

We have calculated the distribution $P(s)$ of adjacent level
spacings $s$, where $s= {(E_{i+1}-E_{i}) } / {<E_{i+1}-E_{i}>}$.
Typical result are presented in Fig. \ref{fig1}. It can be clearly
seen that there is almost no change in the distribution as
function of the disorder once the disorder is above $W=1$.

This unusual behavior of the nearest-neighbor level spacing can be
attributed to the special form of the Cayley-tree. Half of the
sites in the tree are boundary "leaves"- sites at the boundary 
of the tree
which are not connected any further. This peculiar structure of
the tree is known to lead to unusual behavior such as a jump in
the participation ratio at the mobility edge \cite{mirlin94}. In
Ref. \cite{mirlin97} this peculiarity is remedied by connecting
each of the boundary leaves randomly to two other leaf sites.
Thus, the local structure of the Cayley-tree is preserved, while
there are no boundary leaves. $P(s)$ for such a tree is depicted
in Fig. \ref{fig2}. As expected the distribution is shifting from
the Wigner surmise distribution (characteristic of extended
states),
\begin{eqnarray}
   P_W(s)=\frac{\pi}{2}[ s ]\exp[-\frac{\pi}{4}[ s^2]],
\label{wigner}
\end{eqnarray}
to a Poisson distribution (localized states),
\begin{eqnarray}
   P_P(s)=e^{-s} .
\label{poisson}
\end{eqnarray}

\begin{figure}\centering
\epsfxsize5.5cm\epsfbox{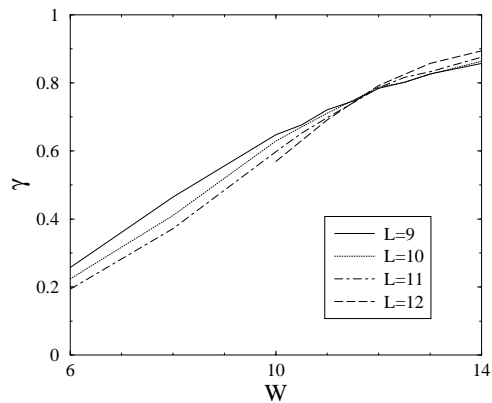} \caption{ $\gamma$ as function
of $W$ for different tree sizes $L$. The typical behavior for
finite size transition is seen, where a crossing in the size
dependence of $\gamma$ between the metallic (small values of $W$)
and localize (large value of $W$) regime is seen.} \label{fig3}
\end{figure}

We can recognize the Anderson transition also by noting that all
curves intersects at $s \sim 2$ and the peak of the distribution
"climbs" along the Poisson curve for larger values of $W$. The
transition point can be established more accurately, as shown in
Ref. \cite{shklov92}, from calculating $\gamma$:
\begin{equation}\label{gamma}
    \gamma = \frac{{\int_{2}^{\infty}P(s)}-{\int_{2}^{\infty}P_{w}(s)}}
    {{\int_{2}^{\infty}P_{p}(s)}-{\int_{2}^{\infty}P_{w}(s)}}.
\end{equation}
$\gamma\rightarrow 0$ as the distribution tends towards the Wigner
distribution, while $\gamma\rightarrow 1$ if the distribution
approaches the Poisson distribution. One expects that as the
system size increases, the finite size corrections will become
small resulting in a distribution closer to Wigner distribution in
the metallic regime and to Poisson in the localized one. At the
transition point the distribution should be independent of the
system size. Indeed, this is the behavior seen in Fig. \ref{fig3}
in which $\gamma$ decreases with system size for small values of
$W$ while it increases with size for large values of $W$. All
curves seem to cross at a particular value of disorder signifying
the critical disorder.

\begin{figure}\centering
\epsfxsize5.5cm\epsfbox{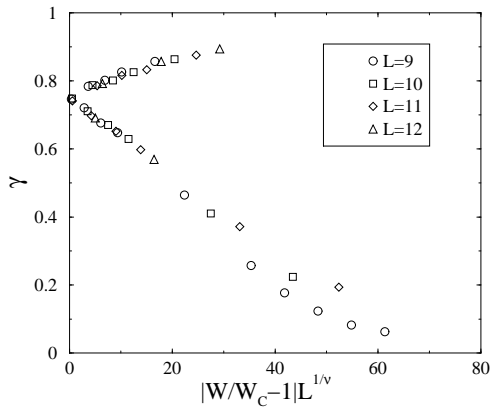} \caption{ The scaling of
$\gamma$ according to Eq. (\ref{scaling}) for different tree sizes
$L$. Two branches, corresponding to the metallic and localized
regimes, appear.} \label{fig4}
\end{figure}

From finite size scaling arguments \cite{shklov92} one expects
that around the critical disorder $\gamma$ will depend on the the
disorder  and tree size in the following way:
\begin{eqnarray}
\gamma(W,L) = \gamma(W_c,L) + C\left|{{W}\over{W_c}}
-1\right|L^{1/\nu} , \label{scaling}
\end{eqnarray}
where $C$ is a constant. This relation enables us to extract the
critical disorder $W_c=11.44^{+0.08}_{-0.04}$ and the critical
index $\nu=0.51^{+0.05}_{-0.04}$. The scaling of the numerical
data according to Eq. (\ref{scaling}) is depicted in Fig.
\ref{fig4}. Two branches corresponding to the metallic and
localized regimes are clearly seen. The critical index $\nu$ fits
rather well the mean field results mentioned above. On the other
hand, the critical disorder for the Cayley-tree is lower than in
the three dimensional case. Thus, while from the extrapolation
equation of $\nu$ one concludes that the dimensionality of the
Cayley-tree is infinity (as expected on geometrical grounds), from
the extrapolation equation of the critical disorder one concludes
that $d=2.7$. This value does not correspond neither to the
geometric dimensionality nor to the spectral dimensionality
$d_s=4/3$.

\begin{figure}\centering
\epsfxsize5.5cm\epsfbox{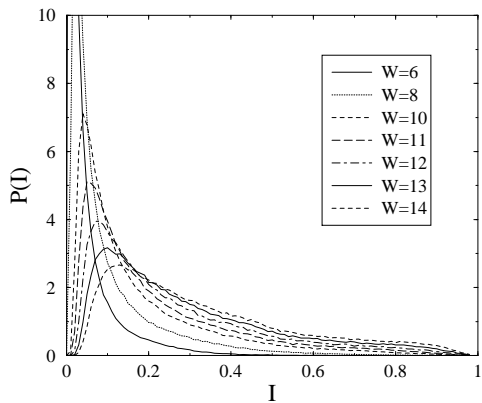} \caption{The distribution of
the IPR for different values of disorder $W$ for an $L=10$ tree.}
\label{fig5}
\end{figure}

It is also interesting to check the behavior of the inverse
participation ratio (IPR) defined as:
\begin{equation}\label{ipr}
I=\sum_r|\psi_i(r)|^4.
\end{equation}
In the metallic regime $I\sim 1/N$, while in the localized regime
$I\sim 1$. The distribution of $I$ as function of the disorder is
depicted in Fig. \ref{fig5}. For small values of $W$ the
distribution is peaked at small values of $I$, as expected in the
metallic regime, while for larger values of $W$ the distribution
is very wide. The transition of the distribution between the
metallic regime and the localized one is, as expected, smooth.

In conclusion, the spectral statistics of a Cayley-tree depends
strongly on the boundary condition. When the boundary leaves are
connected to each other in a way that preserves the local
structure of the tree, a clear Anderson transition is observed. As
expected the critical index $\nu$ corresponds to its mean field
value. On the other hand, the critical disorder for which the
localization transition occurs does not fit into the usual
extrapolation formulas when either the geometric or the spectral 
dimension of the Cayley-tree are used.

We acknowledge very useful discussions with A. D. Mirlin on the importance
of the boundary sites, and support from the Israel Academy of  Science.

\end{document}